# John Archibald Wheeler[1]

## 1911—2008

### *A Biographical Memoir by*
### *Kip S. Thorne*

John Archibald Wheeler was a theoretical physicist who worked on both down-to-earth projects and highly speculative ideas, and always emphasized the importance of experiment and observation, even when speculating wildly.  His research and insights had large impacts on nuclear and particle physics, the design of nuclear weapons, general relativity and relativistic astrophysics, and quantum gravity and quantum information.  But his greatest impacts were through the students, postdocs, and mature physicists whom he educated and inspired.

He was guided by what he called the *principle of radical conservatism*, inspired by Niels Bohr:  base your research on well established physical laws (be conservative), but push them into the most extreme conceivable domains (be radical).  He often pushed far beyond the boundaries of well understood physics, speculating in prescient ways that inspired future generations of physicists.

After completing his PhD with Karl Herzfeld at Johns Hopkins University (1933), Wheeler embarked on a postdoctoral year with Gregory Breit at NYU and another with Niels Bohr in Copenhagen.  He then moved to a three-year assistant professorship at the University of North Carolina (1935-37),  followed by a 40 year professorial career at Princeton University (1937-1976) and then ten years as a professor at the University of Texas, Austin (1976-1987).  He returned to Princeton in retirement but remained actively and intensely engaged with physics right up to his death at age 96.

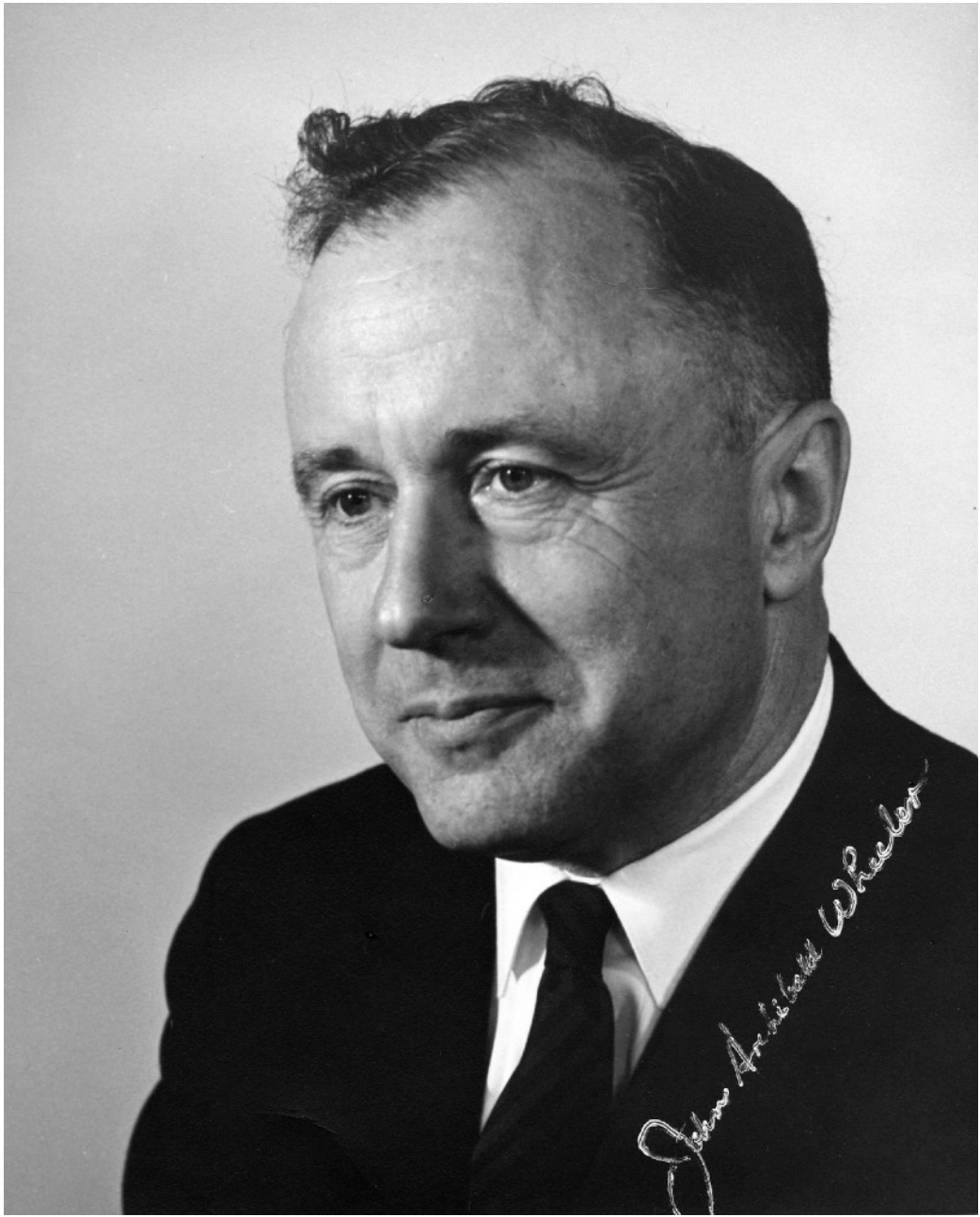

John Archibald Wheeler, ca. 1955.  [Credit: AIP Emilio Segrè Visual Archives]

### The Ethos of John Archibald Wheeler

In demeanor, John Wheeler had an air of formality, so as a student I always called him "Professor Wheeler" — a rather reverential "Professor Wheeler".  The day after I defended my PhD dissertation under his guidance, I telephoned his home and asked his wife Janette if I could speak with Professor Wheeler.  In a kindly voice, she responded, "You have your PhD now, Kip, so you can call him Johnny," and I have done so since then.   In this biographical memoir I'll call him Wheeler, John, or on rare occasion Johnny, depending on the level of formality or personal affection I wish to convey.

Over the decades of our friendship, I came to appreciate and enjoy John's playful side. For example, he loved explosions, though at age ten he had mangled a forefinger and thumb playing with dynamite caps.  In 1971, at a large, formal banquet in the Carlsberg Mansion in Copenhagen, John surreptitiously lit a string of fire crackers and threw it behind his chair to celebrate his 60th birthday.  It caused quite a commotion among the diners, but only I and perhaps one or two others sitting beside him were aware that he was the culprit, and why.  He kept a completely straight face.

John understood the psychological impact that a pithy phrase or the name of a concept could have on researchers and nonscientists alike, so he spent much time lying in a bath of warm water, thinking about possible names and phrases.  Among his coinages are[2]

• *sum over histories* (for Feynman's path-integral formulation of quantum mechanics),
• *moderator* (for the material that slows neutrons in a nuclear reactor),
• *stellarator* (for a plasma magnetic confinement device),
• *wormhole* (for a topological handle in the geometry of curved space),
• *black hole* (for the object left behind when a star implodes),
• *a single quantum cannot be cloned* (for a theorem that limits quantum amplifiers)
• *it from bit* (John's speculation that quantum information is the foundation of all reality)
• *a black hole has no hair* (for uniqueness theorems about black holes).

Regarding *no hair*, Janette once commented to me about Johnny's naughty side.

John was unfailingly polite.  His former student David Sharp gave an example in a 1977 letter to John:  "One day [in the early 1960s when you and I were working together] a man came to see you.  He had a 'theory' of something or other that he wanted to explain.  It became clear after about 30 seconds that the man was a 'crackpot.' … As the discussion dragged on, I began to seethe with impatience. …. But not you.  You treated the man with respect. … You met his ideas head on and quickly but kindly demonstrated the flaws in them. … I'm sure that when the man left he was still convinced of the basic correctness of his 'theory'.  But he did acknowledge the flaws (which were devastating) and I'm equally sure that he felt that he had been treated fairly."

Unfailingly polite?  Well, almost.  On exceedingly rare occasions, when a special need arose, John could be blunt.  Dick Feynman described an example to me in the 1970s, when we were both a bit inebriated at a party:  "When I was his student, Wheeler was sometimes too fast for me," Feynman said.  "One day we were working on a calculation together.  I could not see how he got from this point to the next.  'Little steps for little people,' Wheeler said, as he spelled out for me the omitted steps."  This is the only time I, Kip, *ever* heard *any* former student describe John behaving so impolitely.  I can only speculate 1. that Feynman had been displaying great brashness and arrogance and Wheeler felt he needed to be shown that he was not yet a great master of all physics, and 2. that Wheeler knew Feynman could take such criticism without being seriously damaged.  Evidently, the lesson stuck indelibly; Feynman remembered it with chagrin decades afterward.

In his later years, Wheeler developed a reputation for proposing weird, and seemingly crazy ideas.  One day in 1971 Wheeler, Feynman and I had lunch together at the Burger Continental near Caltech.  Over Armenian food, Wheeler described to Feynman and me his idea that the laws of physics are mutable: Those laws must have come into being in our universe's Big Bang birth, and surely there are other universes, each with its own set of laws. "What principles determine which laws emerge in our universe and which in another?" he asked.

Feynman turned to me and said, "This guy sounds crazy. What people of your generation don't know is that he has always sounded crazy. But when I was his student [30 years earlier], I discovered that, if you take one of his crazy ideas and you unwrap the layers of craziness one after another like lifting the layers off an onion, at the heart of the idea you will often find a powerful kernel of truth." Feynman then recalled Wheeler's 1942 idea that positrons are electrons going backward in time, and the importance of that idea in Feynman's Nobel Prize-winning formulation of quantum electrodynamics.[3]

Today string theorists are struggling to figure out what determines which of the plethora of quantum vacua (and their associated physical laws) in the string theory landscape actually occurred in the birth of our universe, or any other universe. This is a concrete variant of Wheeler's question about what principles determined which laws arose, a variant informed by 47 intervening years of quantum gravity research; and it is an example of Wheeler's prescience — a prescience that is much more appreciated today than in the prime of his career.

John was the principal mentor for roughly 50 PhD dissertations, 50 undergraduate senior theses, and 40 postdoctoral students.[4] His mentoring techniques and effectiveness were remarkable, and so I patterned many of my own techniques after his.

He was tremendously inspirational: In 1962, I had just arrived at Princeton as a graduate student. My dream was to work on relativity with Professor Wheeler, so I knocked on his door with trepidation. He greeted me with a warm smile, ushered me into his office, and began immediately (as though I were an esteemed colleague, not a total novice) to discuss the mysteries of the gravitational collapse of a star at the end of its life. I emerged an hour later, a convert and disciple.[5] Much of my research over the subsequent decade dealt with gravitational collapse, the black holes it produces, and related topics.

John provided detailed guidance for beginning students. Daniel Holz, his last student, wrote in a blog on the day of Wheeler's death:[6] "[In 1990, as an undergraduate looking for a senior thesis project,] I waltzed into Wheeler's office and asked if he had any projects I could work on. I staggered out of his office four hours later, laden with books, a clearly defined project in my hands."

Robert Geroch (a PhD student of Wheeler's in the mid 1960s) has described Wheeler's mentoring style with strong PhD students:[7] "Wheeler had a global view. He forced you to look out and not be too small. 'If you want to know the answer to this,' he would say, 'let's phone Madam Choquet in Paris right now.' 'If you're interested in topic X, then we better fly in Roy Kerr from Texas to explain it to us.' One comes to graduate school with a kind of 'backing off' attitude, an awe of the big names. He was very good at breaking that." Among the colleagues with whom Wheeler put Geroch in touch were Stephen Hawking and Roger Penrose, and as a result, Geroch, as a student, became during that era perhaps the third most influential person after them in applying techniques of differential topology to the study of generic singularities in the structure of spacetime.

Bill Unruh (a Wheeler PhD student in the early 1970s) recalls:[8] "I had just got started working on my first research problem and had a few extremely vague ideas. I mentioned them to Wheeler one day, and he said, 'I've received this invitation to a workshop in Gwatt, Switzerland. Would you like to go and present your results?' I was torn because I didn't have any results to present. And then he said, 'Here, I'll write out this telegram,' and he wrote one saying 'Would you please invite Bill Unruh to give a talk.' He handed it to me and said, 'Please phone this in to the telegraph office.' So I wandered around for two or three hours agonizing over whether to send this telegram, because if I sent it, I was committed. I finally did send it and then had three months to get some results worth presenting."

John was driven by an intense desire to know how Nature works, at the deepest level. In 1932-1952, like most all physicists, he presumed that elementary particles are Nature's most fundamental building blocks, so he focused his research on ***particle physics and nuclear physics***, and in a related detour he devoted great ingenuity and energy to the development of **nuclear weapons**.   In 1952-1976, he focused on curved spacetime, as embodied in Einstein's **general relativity and its quantization**, as Nature's more likely most fundamental building block; and from 1976 onward he focused on **quantum information** as the most likely foundation for all we see.  In the remainder of this biography, I shall describe, in each of these areas, some of John's research and some inspirational ideas that he fed to others.

## Nuclear and Particle Physics[9]

John entered Johns Hopkins University in 1927 at age 16, began research in his third year, and, bypassing the bachelor's degree, continued nonstop to a PhD under Karl Herzfeld in 1933.  In John's PhD thesis, he applied the still rather new quantum theory to the scattering and absorption of light by helium atoms.  During his fifth year at Hopkins, James Chadwick, in England, discovered the neutron, giving birth to nuclear physics.

At NYU in John's first postdoctoral year (1933-34), with his advisor Gregory Breit he calculated the scattering of photons off each other, a process not observed in the laboratory until 63 years later, when intense enough lasers became available. His second postdoctoral year (1934-35), in Copenhagen with Niels Bohr, was largely a period for consolidating his understanding of physics and developing his own viewpoints, very much influenced by Bohr.

This led to nuclear physics as John's prime focus at the University of North Carolina (1935-1937).  In a project with

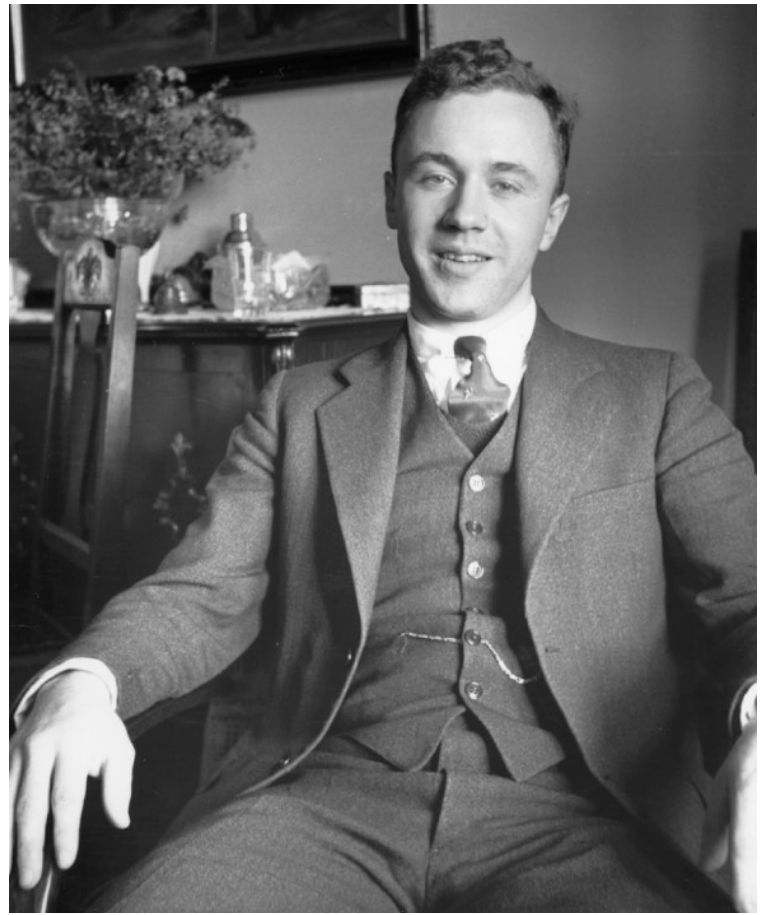

John Wheeler as a postdoc of Niels Bohr  in Copenhagen, 1934.  [Credit:  AIP Emilio Segre Visual Archives, Wheeler Collection.]

Edward Teller (who in Copenhagen had became John's close friend), he pioneered the study of rotational states of nuclei. Alone, he developed a model of nuclei based on "resonating group structure" (later called "clustering") in which the wave function is built up of components that describe the neutrons and protons as clustering into alpha particles and other tightly interacting groups. As a byproduct of this analysis he invented the S-matrix (scattering matrix), which became a major tool for nuclear and particle physics over subsequent decades. In North Carolina he also embarked on mentoring PhD students. His first was Katherine Way, who went on to an eminent career in nuclear physics.

In December 1938, not long after John's move to Princeton, Otto Hahn and Fritz Strassmann in Germany bombarded uranium with slow neutrons, producing nuclear fission (though they did not know what they had; it was Otto Frisch and Lise Meitner who interpreted the data as fission products). Frisch told Bohr, who carried the news on a transatlantic ship to Princeton. There Bohr and Wheeler elaborated George Gamow's liquid drop model of the nucleus and used it to develop the theory of nuclear fission, in one of the most important papers of Wheeler's career. From the Bohr-Wheeler theory it was easily deduced that the ideal nuclei in which to trigger fission via slow-neutron bombardment are uranium 235 (which was used, unknowingly, by Hahn and Fritz) and plutonium 239—which was unknown at the time as it has a small enough half life not to be found in Nature. Plutonium 239 became the foundation for nuclear reactors, in which it is produced artificially in large quantities, and for the first atomic bomb (the Trinity test).

World War II and the atomic bomb effort interrupted much of John's career; see below.

Immediately after the war, impressed by the major discoveries about fundamental particles that had come from cosmic-ray experiments at other institutions (particularly Carl Anderson's lab at Caltech), John proposed, created, and led a cosmic-ray laboratory at Princeton.

He became enamored of the muon (which was cleanly delineated from the pion, experimentally, only in 1947), because the muon's lack of coupling to the strong nuclear force made it much simpler than most other particles. With the aid of observations in his cosmic-ray lab and elsewhere, he gave strong evidence that in all respects except mass the muon's properties are the same as those of an electron.

He focused on atoms in which an electron is replaced by a muon (mu-mesic, later called mu-mesonic, atoms), finding them interesting not only in principle, but because the muon, with its much heavier mass, is more tightly bound to the nucleus than the electron it replaces and so can probe nuclear properties much better. Accordingly, he developed in detail the theory of mu-mesic atoms and linked the theory to experiment, including observations, in his cosmic-ray laboratory by W. K. Chang, of the gamma-ray cascade emitted as the muon in a mu-mesic atom drops from one energy level to another.

In 1949, with his student Jayme Tiomno, John identified the universality of the weak interaction in which neutrons, muons and electrons participate: that the same weak coupling constant governs the beta decay of a neutron (to form a proton, electron, and electron antineutrino); the beta decay of a muon (to form an electron, electron antineutrino, and muon neutrino); and the charge exchange reaction in which a muon is captured by an atomic nucleus and there combines with a proton to form a neutron and muon neutrino. This universality was identified independently by Giampietro Puppi, and a lovely triangle that displays this universality graphically, drawn by Tiomno and Wheeler, but not by Puppi, wound up named the "Puppi triangle."

In the fall of 1949, following shell-model insights of Hans Jensen and Maria Goeppert-Mayer, John realized that in big nuclei, a single nucleon, constrained by liquid-droplet tension, could travel around the rest of the nucleus in a large orbit, deforming the nucleus substantially. He inserted this idea and its quantitative analysis into the manuscript of a paper on a broader topic that he was writing with Niels Bohr and David Hill. Bohr, as was his wont, sat on the paper for many many months, trying to perfect it, and in the meantime, John's idea was discovered independently by James Rainwater at Columbia University and led to Rainwater's sharing a Nobel Prize. Of this, Wheeler has written[10] "…I learned a lesson. When one discovers something significant, it is best to publish it promptly and not wait to incorporate it into some grander scheme. Waiting to assemble all the pieces might be all right for a philosopher, but it is not wise for a physicist." But he did not blame Bohr, for whom he had great affection and reverence; not at all. He just blamed himself.

By the early 1940s, John had formulated and embarked on his quest to understand Nature at its deepest level. His initial hope for the fundamental building block of everything was particles. For a short while he speculated that perhaps, somehow, everything in the universe is made solely from electrons and positrons, but all he ever succeeded in achieving in this direction was the prediction and theory of an almost endless family of short lived "atoms" built from them, which he called "polyelectrons". Of these, the simplest, positronium (one electron and one proton), and the positronium ion (two electrons and one positron) have been created and studied in the laboratory and compared with his theory.

John had greater success in an effort, with Feynman, to remove fields entirely from *classical* electrodynamics, making it a theory based solely on particles. They did this by writing the direct action-at-a-distance Lienard-Weichert force of one charge particle on another as half the retarded force plus half the advanced force, which is time symmetric and leads (i) to no interaction of a particle with itself and thus no infinite self energy to be renormalized, and (ii) to the standard radiation reaction force, which arises, without any radiation field, from the interaction of the accelerated particle with all the other charged particles in the universe (which play the role of *absorbers*). The vision for such a field-free theory came from Feynman; the principal ideas for how to make it really work came from Wheeler, as Feynman describes in great detail in his Nobel Prize lecture.[2]

This field-free classical theory became a major foundation for Feynman's formulation of quantum electrodynamics, but not for Wheeler's dream of a full particles-only formulation of physics. There it was a dead end. A few years after its completion, Wheeler gave up on his particles-only dream and switched to fields-only, in particular, the relativistic gravitational field or spacetime curvature embodied in general relativity. To this I will return after a diversion.

## Nuclear Weapons

Shortly after the Japanese bombing of Pearl Harbor, John plunged full force into the American effort to build an atomic bomb. In January 1942, he joined Arthur Compton's "Metallurgical Laboratory" at the University of Chicago to work on the world's first test reactor, designed to explore the production of plutonium 239 via a nuclear chain reaction. Then in March 1943, Compton assigned him to be the project's liaison scientist for the DuPont Company's project to design and then build the first large-scale plutonium-production reactor at Hanford, Washington. John urged a conservative design that would allow for the possibility that some then unknown atomic nucleus with a very high absorption cross section for slow neutrons might be formed in the fissions, and thereby poison the chain reaction — as indeed did happen.

On October 25, 1944, one month after the poisoning discovery, John's brother Joe was killed in military action in Italy. John, who had been very close to Joe, was devastated. Thereafter he never forgave himself for failing to press to initiate the atomic bomb effort a year or two earlier. That might, he reasoned, have resulted in a much earlier bomb which might have ended the war before Joe and millions of others were killed. This weighed heavily on John for the rest of his life and contributed substantially, I think, to his political conservatism on issues of national defense.

After Joe's death, John doubled down and worked harder than ever on the bomb effort. When the bombs were ultimately dropped on Hiroshima and then Nagasaki with a horrific loss of civilian life that ended the war, John had no misgivings, by contrast with Robert Oppenheimer and many other physicist contributors to the bomb effort.

At the end of the war, John returned to fundamental physics, until the Soviet Union tested its first atomic bomb, in August 1949. The reaction in America was panic, bomb shelters, and atomic bomb drills, even in my little elementary school in rural Utah. The Russian bomb test prompted Teller to urge a crash program to develop the hydrogen bomb (H bomb). Oppenheimer opposed it, John backed it, and President Truman ordered it to go forward. John joined Teller in Los Alamos to work on the bomb's design. A year later, when an innovation by Teller and Stanislaw Ulam made the H bomb seem, for the first time, truly feasible, John set up a satellite bomb design effort at Princeton,[11] to work quasi-independently of the other design effort in Los Alamos (though with frequent communication): a two-track approach also being pursued by Soviet physicists.

John tried, and failed, to get eminent senior physicists to join the design effort, so he assembled a group of graduate students and fresh postdocs to do the work under his guidance. As described by Ken Ford, a member of his team, John "reduced what was known or guessed about reaction rates and the properties of matter in extremis to a set of coupled differential equations of such simplicity that they could be handled numerically on a then-available computer—the National Bureau of Standards SEAC machine—whose total memory capacity was less than 3 kilobytes." John's students and postdocs programmed the computer to model, with these equations, the first planned test of the Teller-Ulam idea. (Their earlier, promising numerical results, achieved on an even more primitive computer, played an important role in June 1951, in convincing the Atomic Energy Commission's General Advisory Committee to recommend moving forward.) In 1952, with the help of SEAC, John's group predicted to within 30% the yield of that first thermonuclear test explosion, code named Mike.

In the Soviet Union, the Teller-Ulam idea was invented independently by Andrei Sakharov and Yakov Borisovich Zel'dovich and led to a Soviet H bomb. A few years later, Wheeler, Sakharov, and Zel'dovich all embarked on research in relativistic astrophysics, and in 1969 I found myself in a hotel room with the three of them, at a relativity conference in Tbilisi, Georgia, USSR.  It was remarkable there to see the camaraderie and deep mutual respect of these three "cold war" physicists for each other.

## General Relativity and Quantum Gravity

In 1952, a few months before the Mike thermonuclear test, John saw his weapons work nearing an end, and so arranged to teach a full year course on relativity.  It was the first relativity course offered at Princeton since 1941 — an indication of the extent to which relativity, in that era of rich nuclear physics, had become a backwater.  John viewed relativity as a subject ripe for exploration and for great discoveries, a subject "too important to be left to the mathematicians".  And maybe, just maybe, curved spacetime would turn out to be the ultimate foundation for everything.  Hence, his eagerness to teach a relativity course:  "If you want to learn, teach" he often said.

Over the next few years, John developed his own, unique viewpoint on relativity: a viewpoint in which the *geometry* of curved spacetime was central, a geometry visualized in pictures of curved surfaces and bending world lines, a geometry that became a foundation for physical intuition.  Charles Misner and I, as John's students, learned that viewpoint from him, and in 1973 we three codified it in our textbook *Gravitation*.  Almost simultaneously, Steven Weinberg codified a field-theoretic viewpoint on general relativity in his textbook *Gravitation and Cosmology*.  Wheeler's geometric viewpoint came to dominate research on classical general relativity, while Weinberg's field-theoretic viewpoint has dominated most modern cosmology research.

By 1956, John had identified a plethora of fascinating research projects in relativity, and his garden of ideas and flowering projects grew rapidly over the subsequent years, as did his entourage of students, postdocs, and senior colleagues.  One can get some sense of the richness of John's ideas from his 1963 lectures at a physics summer school in Les Houches, France.[12]

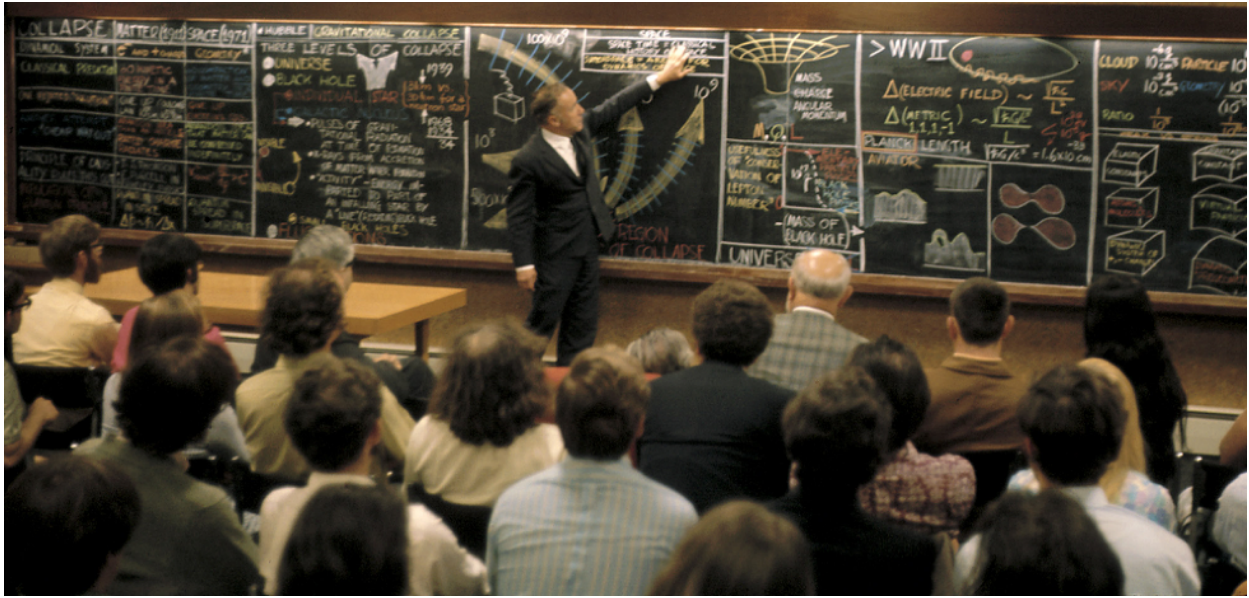

John Wheeler giving an inspirational lecture to colleagues at a conference at the Institute of Astronomy in Cambridge, England, in summer 1971.  John's style was to cover a huge blackboard with colored chalk drawings before the lecture, then work his way through them as the lecture progressed.  [Credit: Kip Thorne]

By the early 1970s, John's Princeton group had grown to about 15 (unusually large for a theory group in those days), and as Bill Unruh recalls, "Wheeler, himself, was the source of the key initial ideas for the research of most everyone in the group."  And by the early 1970s, relativity had become a major branch of physics and was entering a golden age, thanks in considerable measure to the theoretical research of John and his intellectual progeny, and thanks to observational discoveries of quasars, pulsars, and compact X-ray sources (all energized by black holes or neutron stars), and the discovery of the cosmic background radiation from the big bang.

John's interest in relativity was triggered in January 1951, when he studied the 1938-39 work of Robert Oppenheimer and student George Volkoff on neutron stars, and also the work of Oppenheimer and student Hartland Snyder on the collapse of a sufficiently massive star—which (they found) leads the star to "cut itself off from the rest of the universe" and form an infinite-density singularity at its center, i.e. leads it to form what John, seventeen years later, would dub a "black hole".   So it was natural that some of John's earliest projects built on Oppenheimer's work.

With his student Kent Harrison and postdoc Masami Wakano, John asked, "What is the endpoint of thermonuclear evolution for stars of various masses?" They catalogued the entire range of absolute-endpoint objects: a continuous family with increasing central density, from cold white dwarfs made of iron 56 with central densities up to 2.5 x $10^8$ g/cm$^3$, through unstable objects of intermediate densities, to neutron stars with densities 3 x $10^{13}$ to 6x$10^{15}$ g/cm$^3$, and onward into unstable objects with densities increasing toward infinity. This helped solidify the conclusion that sufficiently massive stars (or, as John liked to think of it, stars containing a sufficiently large number of baryons) must undergo the kind of gravitational collapse that Oppenheimer and Snyder had described mathematically.

John was highly skeptical of the Oppenheimer-Snyder conclusions about the collapse. He focused particularly on the singularity (with infinite density and infinite curvature of spacetime) predicted to form deep inside the cut-off sphere (inside what today we call the event horizon). There, he argued, the laws of classical general relativity must break down, and be replaced by laws of quantum gravity that result from "a fiery marriage" of general relativity with quantum theory. This singularity and the *issue of the final state* of massive stars that are prone to collapse to form it, became a major focus of his research and that of his students.

In June 1958, at a Solvay Congress,[13] John rejected the predicted singularity as physically unreasonable and speculated about the collapse's true final state: "… no escape is apparent except to assume that the nucleons at the center of a highly compressed mass [where the singularity is trying to form] must necessarily dissolve away into radiation…at such a rate or in such numbers as to keep the total number of nucleons from exceeding a certain critical number [so the final state can be a neutron star]." Oppenheimer, who was present in the audience, was unpersuaded. And a few years later, with the help of David Sharp, I myself talked John out of including this seemingly outrageous speculation in a book that I was co-authoring with John,[14] though he continued to espouse his speculation elsewhere.

Then, a few years after that, Stephen Hawking discovered *Hawking Radiation* from black holes — a form of radiation very much like John's speculation. When Hawking and John's former postdoctoral student James Hartle devised a derivation of Hawking radiation in which the singularity inside the collapsing star participates in producing the radiation in a manner somewhat similar to John's speculation,[15] I regretted my efforts to suppress John's wild idea, and came to appreciate his prescience.

By 1962, John's entourage had elucidated in crystal clear form what was going on in the Oppenheimer-Snyder calculation: that a horizon forms at the cutoff sphere, hiding the interior from view; and in 1968 John introduced the name *black hole* to describe the resulting object. But for John the issue of the final state *inside* the horizon, the singularity, remained the most important focus, motivating much of his subsequent work on quantum gravity; see below.

In the late 1960s and 1970s, a major focus for John's entourage and others was the physics of black holes. It was particularly important to know whether black holes are stable against small perturbations. For that we all looked back to a pioneering, 1957, stability analysis by John and his student Tullio Regge. In 1957 the horizon was not understood, so it was not fully clear what inner boundary conditions Regge and Wheeler should impose on their equations. Once that was sorted out, in the late 1960s, Charles Misner's student C. V. Vishveshwara was able quickly to complete the Regge-Wheeler analysis and prove that nonrotating black holes are stable. The stability of spinning black holes soon followed, proved by my own students Saul Teukolsky and Bill Press, in an analysis patterned after Regge and Wheeler.

In 1970, Stephen Hawking deduced that, in any process, including highly dynamical ones, the sum of the surface areas of all interacting black holes must increase. Hawking was well aware that this made a black hole's surface area analogous to entropy, but he was highly skeptical that there was any connection. John's graduate student Jacob Bekenstein, by contrast, was quite sure that a black hole's surface area is its entropy in disguise, and he argued vigorously and semiquantitatively for this, with John's strong backing: "It's just crazy enough to be right," John said.  (John liked to quote Gertrude Stein's remark, "It looks strange and it looks strange, and it looks very strange, and then suddenly it does not look strange at all, and you cannot understand what made it look strange in the first place.")  When Hawking used quantum theory to discover that black holes can radiate, he reversed himself, embraced the *Bekenstein entropy* of a black hole, and made understanding it in depth a major focus of his own research.

Among John's entourage in the mid 1950s was a young electrical engineer named Joseph (Joe) Weber, who had recently been hired on the faculty at the University of Maryland, College Park.  John encouraged Joe's interest in relativity and together they explored, mathematically, exact solutions of Einstein's equations for cylindrical gravitational waves, showing that such waves are physical phenomena, not mere figments of the mathematics. (There was much skepticism of their physical reality at that time.) This project with John played a major role in Weber's embarking on his experimental search for gravitational waves from the astrophysical universe, a quest that John strongly encouraged, as he would later encourage me and my LIGO colleagues in our follow-on, and ultimately successful, quest.

Although John's relativity research focused on theory, he paid close attention to observations and experiment.  In 1966, when asked to write a review article on the theory of neutron stars (which had never yet been seen), he chose to include in it a speculation on how they might first be discovered.  "Energy of rotation [of a central neutron star] appears not yet to have been investigated as a source of power [for the Crab nebula]," he wrote.[16]  "Presumably this mechanism can only be effective—if then—when the magnetic field of the residual neutron star is well coupled to the surrounding ion clouds."  This and a similar but more detailed argument by Franco Pacini a year later[17] were the closest anyone ever came, before the 1967 discovery of pulsars, to the correct explanation of what powers the Crab nebula.

Early in his study of general relativity, John became enthusiastic about what he called *geometrodynamics:* the dynamics of the geometry of spacetime, especially in vacuum where there is no matter present to complicate things.

John's first examples of geometrodynamics were toroidal and spherical configurations of electromagnetic waves that are held together (confined) by the gravitational pull and spacetime curvature of the waves' energy. He called these *geons*, and introduced into general relativity a two-lengthscale expansion and self-consistent-field approximation by which to analyze them. John's PhD student Dieter Brill and Brill's undergraduate student James Hartle together used these techniques to analyze *gravitational geons*, in which the electromagnetic waves are replaced by gravitational waves, so the entire geon is an (approximate) solution of the vacuum Einstein equations: vacuum geometrodynamics.

John hoped that fundamental particles such as the proton might turn out to be quantum-gravity analogs of these geons, but he never made any progress in that direction. And his classical geons turned out to be unstable, both to leakage of the waves out of the entity and to a collective radial mode of motion. However, the mathematical techniques that John and then Brill and Hartle introduced to analyze geons, a decade later in the hands of Misner's student Richard Isaacson, produced a rigorous definition of the energy and momentum carried by generic gravitational waves and a rigorous way to analyze the waves' production of and interaction with generic large-scale spacetime curvature. This is one example of the productive chains of influence that flowed from John.

Another example is a clever analysis, in 1957, by John and his student Richard Lindquist, of a vacuum geometrodynamic closed universe, one made of a large number of black holes that interact with each other gravitationally. The dynamics of the universe's expansion and recontraction, Lindquist and Wheeler found, is nearly the same as that of a Friedman model universe that is filled with dust rather than black holes.  The differences, they deduced, become greater as the number of black holes goes down.

To me this is particularly interesting as John's first attempt to explore the motion of a small, strongly gravitating body (the black hole) in a large scale gravitational field (spacetime curvature).  In John's next iteration, with his student Fred Manasse (1963) and with advice from Misner, John introduced matched asymptotic expansions into general relativity to achieve higher rigor and better accuracy, and 20 years later, James Hartle and I (both former Wheeler students), used these same techniques to explore how rotation and non-sphericities of compact bodies (including spinning black holes), when coupled to spacetime curvature, modify the bodies' motion and precession.

One more example of John's chain of influence is numerical relativity.[18] John recognized from the outset that exploring generic geometrodynamics analytically would be exceedingly hard and most likely impossible.  Einstein's equations are too nonlinear.  So — with his team's numerical simulations of the first thermonuclear test explosion recently completed — he urged his entourage to embark on analogous numerical simulations of geometrodynamics.

A 1959 reformulation of Einstein's equations, which split spacetime into space plus time, by Misner together with Richard Arnowitt and Stanley Deser, was ideal for numerical relativity.  This ADM formulation had initial-value (constraint) equations and dynamical equations.  In 1960, Misner solved the constraint equations to obtain a mathematical description of two black holes that are momentarily at rest with respect to each other, and then Lindquist, together with computational scientist Susan Hahn at IBM, solved the evolution equations numerically and thereby watched the black holes fall toward each other.  Unfortunately, the holes' actual collision was beyond the capability of the Hahn-Lindquist computer and code. It was not fully explored numerically until 20 years later, by Larry Smarr and Kenneth Eppley.   Today, numerical relativity in the hands of a younger generation is crucial for analyzing the data from LIGO's gravitational-wave detectors, and is being used to explore generic geometrodynamics,[19] fulfilling John's original, now 60-year-old vision.

John and his entourage were driven into early explorations of *quantum gravity* by both the issue of the final state (the singularity inside black holes), and John's fixation on the deepest foundations of physics.

John's first venture into quantum gravity was a paper he published in 1957 titled "Quantum Geometrodynamics",[20] in which he made educated guesses as to what physical phenomena might result from the fiery marriage of general relativity with quantum theory. Most importantly, he identified the *Planck length* as the characteristic length scale for quantum gravity effects, and he argued that on this length scale space should exhibit quantum foam: a foam of randomly fluctuating curvature and topology, including microscopic wormholes — handles in the structure of space first described classically by Hermann Weyl in 1924, and explored in depth by John and his entourage in the 1950s and early 1960s.

In the early 1960s, when I was John's graduate student, Bryce DeWitt often visited Princeton from North Carolina, for long discussions with John about quantum gravity. I sat in on these discussions, only half understanding, as much went over my head. The discussions led to their formulating together the basic ideas for a quantum theory of gravity: a wave function defined on a superspace of spacelike, 3-dimensional geometries; and an equation — since named the *Wheeler-DeWitt equation* — that governed that wave function. DeWitt took off from there in developing the theory in great detail.[21] Hawking and others have used this theory in their searches for insights into quantum gravity; but, of course, it is only one among several approaches to quantum gravity that look promising today.

## Quantum Information

With his move to Texas in 1976, John transitioned from relativity as his central focus to the role of measurement in quantum physics — an ancient subject on which Bohr and Einstein had jousted in the 1930s, but one which had become an extreme backwater by 1976. John's interest in quantum measurement dated back to his 1934-35 postdoctoral year with Bohr, but only upon arriving at Texas did he plunge into it.

John assembled around himself in Texas an entourage of students, postdocs, and faculty members much like his previous relativity entourage at Princeton. He jump started his entourage with a two-year-long course on quantum measurement, in which they studied a wide range of writings of previous generations. Wojciech Zurek, a graduate student in the entourage, recalls that "The class … often turned into a seminar where visitors and students reported their research or interesting new papers." A collection of readings from that class, published by Wheeler and Zurek,[22] became a resource for others, as Wheeler's group and colleagues elsewhere began to revitalize research on quantum measurement and related areas of quantum physics, transforming them into the modern field of *quantum information.*

Zurek, in retrospect, assesses John's influence thus:[23] "Looking back on Wheeler's ten years at Texas, many quantum information scientists now regard him, along with IBM's Rolf Landauer, as a grandfather of their field. That, however, was not because Wheeler produced seminal research papers on quantum information. He did not—with one major exception, his delayed-choice experiment … . Rather, his role was to inspire by asking deep questions from a radical conservative viewpoint and, through his questions, to stimulate others' research and discovery."

Among the members of John's entourage whom he so stimulated were Zurek and graduate student William Wooters who, in 1982, under John's influence, formulated and proved the theorem that an unknown (unmeasured) quantum state cannot be cloned. Another was postdoc David Deutsch, who, after moving to Oxford formulated and proved in 1985 the possibility of a universal quantum computer (a quantum Turing machine): one that can simulate any other quantum computer with at most a polynomial slowdown. A fourth was Texas assistant professor (1979-1985) Jeff Kimble, who carried out fundamental experiments in quantum optics that produced and measured new, nonclassical states of light, such as photon anti-bunched states and squeezed states.  Later, at Caltech, he made crucial contributions to quantum nondemolition technology for LIGO's gravitational wave detectors.

John's views on quantum measurement were an elaboration of those of Bohr, as embodied in the Copenhagen interpretation of quantum mechanics.  The central facet of an (ideal) quantum measurement, John maintained, is the "collapse of uncertainty into certainty", as embodied in the collapse of the wave function.  He probed this collapse conceptually with his *delayed choice experiment,* a thought experiment in which the experimenter's choice of what to measure can be regarded as influencing the past history of the measured system—and even converting it from being uncertain in the quantum sense to being definite in the classical sense.

John began with a standard "Mach-Zender" interference experiment, though with single photons:  The wave-packet quantum state of a single photon is split in two by one beam splitter and then recombined by another, and then the photon is detected (*measured*) by a photodetector at one or the other output port of the second splitter. If the path lengths between splitters are equal and the second splitter is present, then interference of the recombining wave packets causes every photon to be detected at just one output port.  We know, then, that the photon, emerging from the first splitter, went down both paths and interfered to make one output port always light up and the other always remain dark.  If, on the other hand, the second splitter is absent, then the measured photons wind up equally distributed between the two ports, telling us that each of the photons made a random choice of which path to go down, and went down solely that one: the unique path that led it to the output port where it was detected. The choice of what to measure (of whether to include the second splitter or not) determined which path(s) the photon followed: both, or just one.

John turned this into a "delayed choice" experiment by (conceptually) inserting or removing the second splitter *after* the wave packet passed through the first splitter. The choice of measurement (second splitter or no second splitter) then could be regarded as reaching into the past and making definite which path(s) the photon followed: one or both.  [Several years after John conceived this thought experiment, William Wickes, Caroll Alley, and Oleg  Jakubowicz, actually carried it out at the University of Maryland, getting precisely the result that John knew they would.[24]]

This thought experiment led John to speculate that the universe might be "a self-excited circuit" — a system whose existence and history are determined by measurements, many of them made long after it came into existence.  (John hastened to add that measurements in this, and in Bohr's, sense do not require intelligent life. Each measurement "is an irreversible act in which uncertainty collapses to certainty …, some event in the classical world [such as] the click of a counter, the activation of an optic nerve in someone's eye or just the coalescence of a glob of matter triggered by a quantum event.")

John's self-excited-circuit idea in turn led him to speculate that *information theory is the basis of existence:*  "Trying to wrap my brain around this idea, …, I came up with the phrase 'it from bit'.  The universe and all that it contains ('it') may arise from the myriad yes-no choices of measurement (the 'bits') [that occur during the life of the universe]."

As crazy as this may sound, many quantum information scientists think it respectable. In John's famous words, it just might be "crazy enough to be right."

## Family

John describes the first time he noticed Janette Hegner, at a dance in Baltimore in spring 1933:[9] "She looked me straight in the eye.  No fluttering eyelashes for Janette. … I was attracted to her quick wit, her obvious intelligence, and her commonsense approach to matters we talked about."   Later, after just three dates, they became engaged, but delayed marriage until John returned from his postdoctoral year with Bohr in Copenhagen.

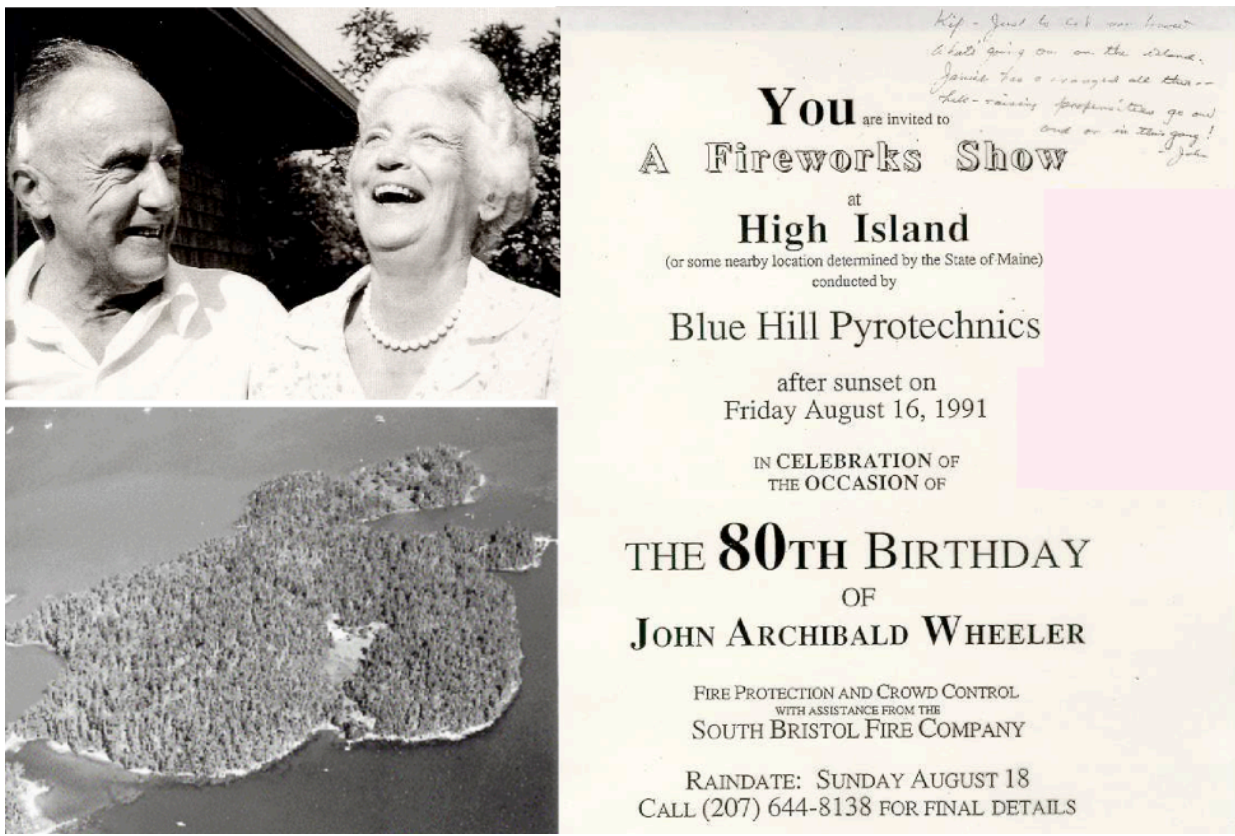

Upper left: John and Janette in 1984 [©1984 Beverly White Spicer].  Lower left: High Island [credit: photo by Jack Lane, courtesy James Wheeler].  Right: Invitation to John's 80th birthday celebration on High Island, with fireworks.

Their marriage was a true partnership, as is clear from John's autobiography,[9] and it lasted robustly for the rest of their lives.  Janette's influence on John was profound, as was hers on him.  Most evenings, in bed, they would read to each other from a book they had jointly chosen.  And together they provided a welcoming, warm home environment for students and visiting physicists, with wonderful meals cooked by Janette.

John, Janette and their three children, Letitia, James and Alison, were a tight-knit, traditional family.  John escaped from Princeton frequently during his career, to far-off places where he could hide and think or interact fruitfully with colleagues, and his family often went with him — for example, seven months in France in 1949–1950; and nine months in Leiden, Netherlands in 1956.  Those sojourns were crucial opportunities for John to develop fresh viewpoints and directions in research, or sometimes simply to complete long-delayed projects.

In 1957, John and Janette bought half of High Island, a 66 acre island in Maine connected to the mainland by a causeway and road.   Thereafter they spent most summers there, with physics students and colleagues visiting frequently for discussions or collaborative work.  Janette and John welcomed my wife Linda and me, and our baby daughter Kares, to stay in a cottage on their island for much of the summer of 1964, as John and I wrote a thin little book on *Gravitation Theory and Gravitational Collapse*.  Janette and John were wonderful hosts. Five years later, Charlie Misner and his wife Susanne had built a home on the Maine coast near High Island, so Linda and I and our two children rented a nearby cottage for a summer of intense writing on our textbook Gravitation — a collaboration that Johnny, Charlie and I treasured.  When the book was finished, John gave Janette, Susanne, and Linda each a gorgeous, large, silver and turquoise pin with an icon of High Island on it, as a memento of their contribution to our idyllic summer months together, with Johnny, Charlie, and me largely sequestered and writing.

In the 1990s and 2000s, with John in (supposed) retirement, he and Janette continued their summer stays in High Island. The rest of the year they lived in a suburb of Princeton, where John continued to go into the office frequently, to interact with physicist colleagues and students.

### Acknowledgments

In writing this biographical memoir, I have relied heavily on articles about John in *Physics Today* published soon after his death, that were written by Ken Ford,[8] by Terry Christensen,[3] and by Charles Misner, Wojciech Zurek, and me,[1] and on materials that the five of us collected in preparation for that writing, and on John Wheeler's autobiography[9], co-authored with Ken Ford.  I thank Ken, Terry, Charles and Wojciech for their large contributions to this memoir, via these materials.  And I also thank Ken, and John's son and daughter James Wheeler and Alison Lahnston for helpful comments and suggestions on this memoir.

## Notes